\documentclass[12pt]{iopart}
\pdfoutput=1
\usepackage{bm}
\usepackage{amssymb}% http://ctan.org/pkg/amssymb
\usepackage{graphicx}
\usepackage{datetime}
\def\d{{\mathrm{d}}}

\def\O{{\mathcal{O}}}

\makeatletter
\graphicspath{{.}{./Figs/}{./pdf/}}
\begin{document}
%------------------------------------------------------------------------------------------------------------------------------------------

\title[``Twisted'' black holes are unphysical.]{
``Twisted'' black holes are unphysical.
}
\author{Finnian Gray, Jessica Santiago, Sebastian Schuster, \\ {\sf and} Matt~Visser}
\address{School of Mathematics and Statistics,
Victoria University of Wellington; \\
PO Box 600, Wellington 6140, New Zealand.}
%------------------------------------------------------------------------------------------------------------------------------------------
%------------------------------------------------------------------------------------------------------------------------------------------
%------------------------------------------------------------------------------------------------------------------------------------------
%------------------------------------------------------------------------------------------------------------------------------------------%------------------------------------------------------------------------------------------------------------------------------------------
%------------------------------------------------------------------------------------------------------------------------------------------
\begin{abstract}\\
So-called ``twisted'' black holes have recently been proposed by Zhang (1609.09721 [gr-qc]), and further considered by Chen and Jing (1610.00886 [gr-qc]), and more recently by Ong (1610.05757 [gr-qc]).  While these spacetimes are certainly Ricci-flat, and so mathematically satisfy the vacuum Einstein equations, they are also merely minor variants on Taub--NUT spacetimes. Consequently they exhibit several unphysical features that make them quite unreasonable as realistic astrophysical objects. Specifically, these ``twisted'' black holes are not (globally) asymptotically flat. Furthermore, 
they contain closed timelike curves that are \emph{not} hidden behind any event horizon --- the most obvious of these closed timelike curves are small azimuthal circles around the rotation axis, but the effect is more general. The entire region outside the horizon is infested with closed timelike curves. 


\medskip
\noindent
D{\sc{ate}}:  19 October 2016; \LaTeX-ed \today {} --- \currenttime
\end{abstract}

%------------------------------------------------------------------------------------------------------------------------------------------
%------------------------------------------------------------------------------------------------------------------------------------------
%\begin{document}
%------------------------------------------------------------------------------------------------------------------------------------------
%------------------------------------------------------------------------------------------------------------------------------------------
\maketitle
%------------------------------------------------------------------------------------------------------------------------------------------
%------------------------------------------------------------------------------------------------------------------------------------------
\clearpage
%------------------------------------------------------------------------------------------------------------------------------------------
%------------------------------------------------------------------------------------------------------------------------------------------
\section{Introduction}
%------------------------------------------------------------------------------------------------------------------------------------------
%------------------------------------------------------------------------------------------------------------------------------------------
\def\d{{\mathrm{d}}}

The ``twisted'' black holes recently proposed by Zhang~\cite{Zhang}, and further considered by Chen and Jing~\cite{Chen}, and by Ong~\cite{Ong}, are certainly Ricci-flat, and so mathematically satisfy the vacuum Einstein equations.
However, these ``twisted'' black holes are merely minor variants on the well-known Taub--NUT spacetimes~\cite{Taub,NUT,Podolsky}. In particular, since these ``twisted'' black holes do not fall into the standard Kerr--Newman class, they certainly violate the \emph{spirit} of the black hole uniqueness theorems~\cite{Robinson, Wiltshire}. Therefore, even without doing any specific calculation,  it is clear that there must be something odd/unusual about these ``twisted'' black holes, something that violates the input assumptions going into the black hole uniqueness theorems, and that this something odd/unusual is likely to render these ``twisted'' black holes unphysical in any realistic astrophysical setting.

In the interests of clarity we shall perform a brief elementary analysis of some of the key features of these ``twisted'' black holes, somewhat along the lines of~\cite{Visser:2007} for the Kerr geometry, relating these ``twisted'' black holes to the more usual Taub--NUT discussion~\cite{Taub, NUT, Podolsky}. 
Specifically, we shall demonstrate that:
\begin{itemize}
\item These ``twisted'' black holes are not globally asymptotically flat.
\item These ``twisted'' black holes actually connect two universes via a timelike wormhole.
\item These ``twisted'' black holes are causally diseased: The entire domain of outer communication is ``totally vicious''~\cite{Minguzzi:2006}.
(Meaning the entire region \emph{outside} the event horizon is infested with closed timelike curves.)
\end{itemize}

To verify these assertions it is sufficient to look at the massless case, the physically problematic issues persist and if anything are actually worse in the massive case.

%------------------------------------------------------------------------------------------------------------------------------------------
%------------------------------------------------------------------------------------------------------------------------------------------
\section{Massless case}
%------------------------------------------------------------------------------------------------------------------------------------------
%------------------------------------------------------------------------------------------------------------------------------------------
\def\d{{\mathrm{d}}}

Consider the spacetime geometry~\cite{Zhang,Chen}
\begin{equation}
\fl
\d s^2 = - \left( r^2-a^2\over r^2+a^2\right) (\d t - 2 a \cos\theta\; \d \phi)^2 
+ \left( r^2+a^2\over r^2-a^2\right) \d r^2 
+ (r^2+a^2) (\d\theta^2 +\sin^2\theta\; \d \phi^2).\quad
\end{equation}
This manifold is Ricci-flat, $R_{ab}=0$,
so that it satisfies the vacuum Einstein equations, and is not Riemann-flat, $R_{abcd}\neq 0$, (so it really is a curved spacetime). In particular the quadratic curvature invariant is not identically zero:
\begin{equation}
R^{abcd} R_{abcd} = 
- {48 a^2(r^2-a^2)([r^2+a^2]^2-[4ar]^2)\over (r^2+a^2)^6} .
\end{equation}
This spacetime is the $m=0$ sub-case of the Taub--NUT variant given in equation (12.1) of reference~\cite{Podolsky}. 

(The Taub--NUT variant given in equation (12.3) of reference~\cite{Podolsky} is physically different --- these two Taub--NUT variants differ not only by a coordinate transformation, but also by a topological identification of some of the points in the manifold.)

%------------------------------------------------------------------------------------------------------------------------------------------
%------------------------------------------------------------------------------------------------------------------------------------------
\subsection{Lack of asymptotic flatness}
%------------------------------------------------------------------------------------------------------------------------------------------
%------------------------------------------------------------------------------------------------------------------------------------------
Let $r \gg a$ and keep only the dominant terms in the metric. Then
\begin{equation}
\d s^2 \approx -  (\d t - 2 a \cos\theta\; \d \phi)^2  + \d r^2 + r^2 (\d\theta^2 +\sin^2\theta\; \d \phi^2).
\end{equation}
This is not globally asymptotically flat, the ``twisting'' encoded in the 
\begin{equation}
- (\d t - 2 a \cos\theta\; \d \phi)^2
\end{equation}
term remains significant no matter how large $r$ gets. 
Note that in this asymptotic regime
\begin{equation}
g_{\phi\phi} \approx r^2\sin^2\theta - 4 a^2 \cos^2 \theta,
\end{equation}
and that this quantity becomes negative for
\begin{equation}
\tan^2\theta \lesssim {4a^2\over r^2}; \qquad \hbox{that is} \qquad r \;|\!\tan\theta| \lesssim 2a.
\end{equation}
In view of the asymptotic condition, this can be further approximated as either
\begin{equation}
r \; \theta \lesssim 2a,  \qquad\hbox{or}\qquad  r\; (\pi-\theta) \lesssim 2a,
\end{equation}
depending on whether one is close to the north-pole or south-pole axis of rotation.
This observation implies that (in this asymptotic region) small azimuthal circles of constant $(t,r,\theta)$ and $\phi\in(0,2\pi)$, encircling  the axis of rotation and of radius $ \lesssim 2a$,  will be closed timelike curves. We shall subsequently see that the existence of closed timelike curves extends all the way down to the horizon, and in fact ultimately contaminates the entire region outside the horizon.

%------------------------------------------------------------------------------------------------------------------------------------------
%------------------------------------------------------------------------------------------------------------------------------------------
\subsection{Two universes connected by a ``bridge''}
%------------------------------------------------------------------------------------------------------------------------------------------
%------------------------------------------------------------------------------------------------------------------------------------------

Another interesting feature of these ``twisted'' black holes~\cite{Zhang,Chen} is that $r=0$ is not the ``point'' at the centre of the spacetime. 
Instead $r=0$ corresponds to a 2-surface of finite surface area, and the $r$ coordinate can be extended to arbitrary negative values. 
To see this, note the following:
\begin{itemize}
\item The metric is invariant under $+r \longleftrightarrow -r$. 
\item At $r=0$ the curvature invariant $R^{abcd} R_{abcd} \to {48\over a^4}$ is finite. (In fact, in a suitable tetrad basis all the tetrad components of the Riemann tensor remain finite at $r=0$.)
\item At fixed $r$ and $t$ the induced 2-metric is
\begin{equation}
\fl
(\d s^2)_2 =  - \left( r^2-a^2\over r^2+a^2\right) (4 a^2 \cos^2\theta\; \d \phi^2 )
+ (r^2+a^2) (\d\theta^2 +\sin^2\theta\; \d \phi^2).\quad
\end{equation}
This is manifestly invariant under $+r \longleftrightarrow -r$, and in particular at $r=0$ we have
 \begin{equation}
(\d s^2)_2 \to  4 a^2 \cos^2\theta\; \d \phi^2  + a^2 (\d\theta^2 +\sin^2\theta\; \d \phi^2).
\end{equation}
Thus the ``point'' $r=0$ is actually a topological 2-sphere with finite surface area:
\begin{eqnarray}
A_0 &=& 2\pi a^2 \int_0^\pi
\sqrt{ \sin^2\theta + 4 \cos^2\theta} \; \d\theta   =  4\pi a^2 \int_0^\pi
\sqrt{ 1 - {\textstyle{3\over 4}} \; \sin^2\theta} \; \d\theta
\nonumber\\
 &=& 4\pi a^2 \times 2\; \mathrm{EllipticE}\left({\textstyle{\sqrt{3}\over2}}\right).
\end{eqnarray}
\end{itemize}
Combining the above, we see that  negative values of $r$ make just as much sense as positive values of $r$.
That is, the natural range of the $r$ coordinate is $r\in(-\infty,+\infty)$.

By considering the zeros of
\begin{equation}
g^{rr} =  \left( r^2-a^2\over r^2+a^2\right),
\end{equation}
we see that there are \emph{two} horizons, located at $(r_H)_\pm=\pm a$. (The regions ``outside'' the two horizons are often called the NUT regions~\cite{Podolsky}.)
By considering the zeros of
\begin{equation}
g_{tt} =  -\left( r^2-a^2\over r^2+a^2\right),
\end{equation}
we see that there are \emph{two} ergosurfaces, located at $(r_E)_\pm=\pm a$. (So the ergosurfaces coincide with the horizon and the ergoregion is empty). The second horizon cannot summarily be ignored the way it is in~\cite{Zhang, Chen,  Ong}. 

The ``bridge'' between the two horizons, $(r_H)_+$ and $(r_H)_-$, which connects the two universes has some vague similarities to a Tolman wormhole~\cite{Hochberg:1998, Molina-Paris:1999}, and some vague similarities to an Einstein--Rosen bridge (non-traversable wormhole)~\cite{Visser:1995}. However it is more commonly referred to as a Taub cosmology; though perhaps Taub wormhole would be better terminology. 

%------------------------------------------------------------------------------------------------------------------------------------------
%------------------------------------------------------------------------------------------------------------------------------------------
\subsection{Taub wormhole?}
%------------------------------------------------------------------------------------------------------------------------------------------
%------------------------------------------------------------------------------------------------------------------------------------------

 In the region $r\in(-a,+a)$ the $r$ coordinate is timelike, while the $t$ coordinate is spacelike. 
To really drive this point home, let us restrict attention to the region $r\in(-a,+a)$, and simply re-label $r\longleftrightarrow t$. Then the new $t$ coordinate has $t\in (-a,+a)$ and the metric becomes
\begin{equation}
\fl
\d s^2 = - \left( a^2+t^2\over a^2-t^2\right) \d t^2  + \left(a^2- t^2\over a^2+t^2\right) (\d r - 2 a \cos\theta\; \d \phi)^2 
+ (a^2+t^2) (\d\theta^2 +\sin^2\theta\; \d \phi^2).\quad
\end{equation}
Viewed in this way, with these new coordinates the region $t\in(-a,+a)$ is an anisotropic cosmology. 
Indeed, since both Ricci and Einstein tensors are zero, this is automatically a homogeneous anisotropic cosmology, and so must fall into the Bianchi classification. In fact it is Bianchi type IX, and is Taub's original cosmological solution~\cite{Taub}. 
It represents a universe undergoing a ``bounce'' in the $\theta$--$\phi$ directions, but a moment of maximum expansion in the $r$ direction. For this reason we shall call it a Taub wormhole. 

Indeed, the maximum analytic extension is a double tower of asymptotic NUT regions, connected by multiple Taub wormholes, similar to the maximum analytic extension of the Reissner--Nordstr\"om or Kerr spacetimes~\cite{Podolsky}.

%------------------------------------------------------------------------------------------------------------------------------------------
%------------------------------------------------------------------------------------------------------------------------------------------
\subsection{Causal structure --- closed timelike curves near the rotation axis}
%------------------------------------------------------------------------------------------------------------------------------------------
%------------------------------------------------------------------------------------------------------------------------------------------

Perhaps the most physically distressing feature of these ``twisted'' black holes is the causal structure.
(See reference~\cite{Ong}, or more generally~\cite{Podolsky}, but note that we are much less sanguine concerning this issue.)

The exact formula for $g_{\phi\phi}$ is
\begin{equation}
g_{\phi\phi} = {4a^2(r^2-a^2) + (r^4+6a^2r^2-3a^4)\sin^2\theta \over r^2+a^2}.
\end{equation}
As per the previous discussion for the asymptotic limit, the curves $\phi\in(0,2\pi)$ which encircle the rotation axis at fixed $(t,r,\theta)$ will be closed timelike curves whenever $g_{\phi\phi} <0$. The boundary of this specific region of closed timelike curves occurs for the closed null curves located by setting $g_{\phi\phi}=0$, which happens when 
\begin{equation}
\sin^2\theta = {4a^2(r^2-a^2)\over r^4+6a^2r^2-3a^4}
= {4a^2(r^2-a^2)\over (r^2+a^2)^2 + 4a^2(r^2-a^2)}.
\end{equation}
This condition can also be recast as 
%\begin{equation}
%\cos^2\theta =  {(r^2+a^2)^2\over (r^2+a^2)^2 + 4a^2(r^2-a^2)},
%\end{equation}
%or even
\begin{equation}
\tan^2\theta = {4a^2(r^2-a^2)\over (r^2+a^2)^2}.
\end{equation}
Note that $\sin^2\theta\in(0,1)$  for $|r|>a$, all the way to $|r|=\infty$. So in ``our'' universe there are two ``cigars'' of these azimuthal closed timelike curves stretching along the rotation axis all the way from spatial infinity to the horizon.  Additionally, there are two more ``cigars'' of these azimuthal closed timelike curves in the ``other'' universe $r\in(-\infty,-a)$.

%------------------------------------------------------------------------------------------------------------------------------------------
%------------------------------------------------------------------------------------------------------------------------------------------
\subsection{Causal structure --- an infestation of closed timelike curves}
%------------------------------------------------------------------------------------------------------------------------------------------
%------------------------------------------------------------------------------------------------------------------------------------------

To see that these ``cigars'' of azimuthal closed timelike curves imply additional causal problems throughout the entire domain of outer communication, let us temporarily work at fixed $r$ and $\theta$. The induced 2-metric on the $t$-$\phi$ cylinder is
\begin{equation}
(\d s^2)_2 = - \left( r^2-a^2\over r^2+a^2\right) (\d t - 2 a \cos\theta\; \d \phi)^2 
+ (r^2+a^2) \sin^2\theta\; \d \phi^2.
\end{equation}
The condition for a curve $t(\phi)$ to be timelike is then 
\begin{equation}
 \left( r^2-a^2\over r^2+a^2\right) \left({\d t\over\d \phi}- 2 a \cos\theta\right)^2  >  (r^2+a^2) \sin^2\theta. 
\end{equation}
Thus for timelike curves
\begin{equation}
\fl
\left({\d t\over\d \phi}\right)\in 
\left(-\infty, 2a\cos\theta - {(r^2+a^2)\sin\theta\over \sqrt{r^2-a^2}}\right) \cup 
\left(2a\cos\theta + {(r^2+a^2)\sin\theta\over \sqrt{r^2-a^2}},+\infty\right).
\label{E19}
\end{equation}
For fixed $\theta$ at large $r$ we have 
\begin{equation}
\left({\d t\over\d \phi}\right)\in 
\left(-\infty, - r\sin\theta\right) \cup 
\left(+r\sin\theta,+\infty\right).
\end{equation}
indicating the usual timelike cone that would be expected for a flat space. 

On the other hand, for fixed $r$ (in either NUT region) and $\theta$ approaching the axis of rotation (either $\theta\to0$ or $\theta\to \pi$),\ we can argue as follows:
There will come a stage where either one or the other of the two intervals in equation (\ref{E19}), for which $\d t/\d\phi$ corresponds to a timelike curve,  will cross the origin ($\d t/\d\phi=0$) and slop over a little into the ``wrong sign'' region.

\begin{figure}[htbp]
\begin{center}
\includegraphics[scale=0.85]{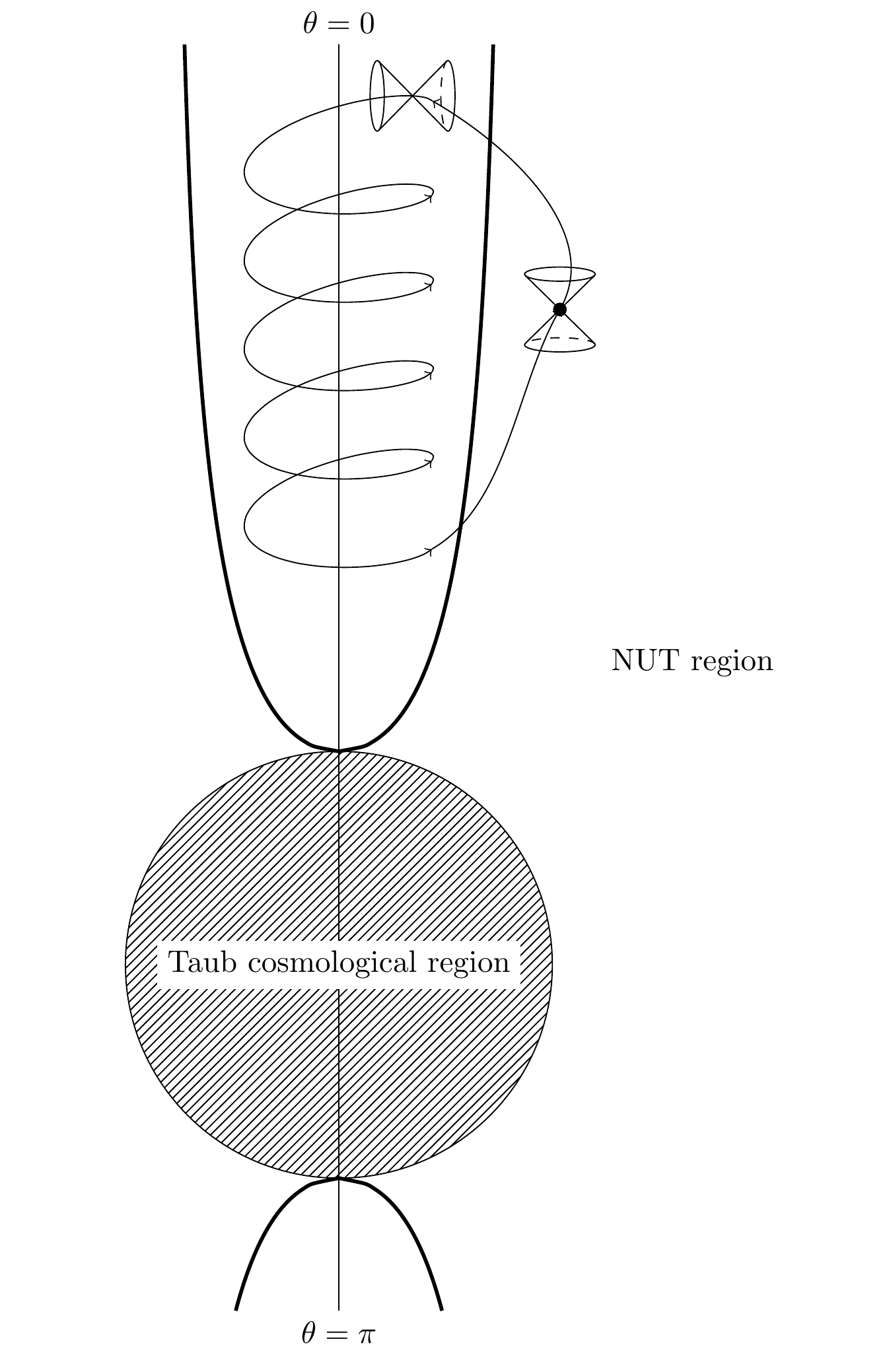}
\caption{Construction of closed timelike curves that cover the entire NUT region. The key observation is that the timelike helices in the ``cigar'' region are not safely hidden behind any sort of horizon.}
\label{F:1}
\end{center}
\end{figure}

This indicates that the timelike cones have ``tipped over'' to such an extent that it is now possible to choose a timelike trajectory that is initially future pointing, but such that after entering the ``tipped over'' region one can choose to follow a helical path that goes backward in the $t$ coordinate. By circling the rotation axis a sufficient number of times one can go arbitrarily far back in the $t$ coordinate. One can then leave the ``tipped over'' region, and return to one's initial starting point, producing a closed timelike curve. (See figure~\ref{F:1}.) This can be done as long as one is initially outside the horizon (in one of the domains of outer communication, one of the NUT regions) and never crosses the horizon --- so the entire domain of outer communication is infested with closed timelike curves. That is, the entire domain of outer communication is ``totally vicious''. 

The critical ``tipping point'' for the timelike cones is seen to occur at
\begin{equation}
\tan\theta = \pm {2a \sqrt{r^2-a^2}\over r^2+a^2},
\end{equation}
which is of course just a rephrasing of the condition $g_{\phi\phi}=0$, which was used to define the ``cigars'' of most manifestly obvious closed timelike curves. 

%------------------------------------------------------------------------------------------------------------------------------------------
%------------------------------------------------------------------------------------------------------------------------------------------
\subsection{Angle surfeit along the axis of rotation}
%------------------------------------------------------------------------------------------------------------------------------------------
%------------------------------------------------------------------------------------------------------------------------------------------
\def\C{{\mathcal C}}
\def\R{{\mathcal R}}

To calculate the angle surfeit along the axis of rotation, note that the circumference and radius of a small disk centred on the axis of rotation are (to sufficient accuracy)
\begin{equation}
\C = 2 \pi \; \sqrt{g_{\phi\phi}};  \qquad   \R = \sqrt{g_{\theta\theta}} \; \theta = \sqrt{r^2+a^2} \; \theta.
\end{equation}
The angle surfeit on the axis is then
\[
\Delta \Phi = \lim_{\theta\to0} \left(\C\over \R\right) - 2\pi 
= 2\pi \left\{\lim_{\theta\to 0}\left({\sqrt{g_{\phi\phi}}\over  \sqrt{r^2+a^2} \; \theta}\right) - 1 \right\} \to \infty.
\]
So in this precise technical sense there is infinite excess angle along the axis of rotation; this is yet another indication of the lack of global asymptotic flatness.

%------------------------------------------------------------------------------------------------------------------------------------------
%------------------------------------------------------------------------------------------------------------------------------------------
\subsection{Asymptotic mass and geodesic (coordinate) acceleration}
%------------------------------------------------------------------------------------------------------------------------------------------
%------------------------------------------------------------------------------------------------------------------------------------------
\def\O{{\mathcal{O}}}
We note
\begin{equation}
g_{tt} = -1 + {2a^2\over r^2} + \O(1/r^4),
\end{equation}
indicating that the mass of the spacetime is (in any meaningful sense) zero. Furthermore the $\Gamma^r{}_{tt}$ Christoffel symbol is
\begin{equation}
\Gamma^r{}_{tt} = {2a^2 r (r^2-a^2)\over (r^2+a^2)^3 } \approx {2a^2\over r^3}, 
\end{equation} 
so the coordinate acceleration of a test particle dropped at rest is inverse cube rather than inverse square. 
(This will change in the next section once we add a mass term.) Note that the coordinate acceleration is always towards the nearest horizon.

%------------------------------------------------------------------------------------------------------------------------------------------
%------------------------------------------------------------------------------------------------------------------------------------------
\section{Massive case}
%------------------------------------------------------------------------------------------------------------------------------------------
%------------------------------------------------------------------------------------------------------------------------------------------
\def\d{{\mathrm{d}}}

In the massive case one introduces an extra parameter $m$ and considers the spacetime
\begin{eqnarray}
\fl
\d s^2 &=& - \left( r^2-2mr-a^2\over r^2+a^2\right) (\d t - 2 a \cos\theta\; \d \phi)^2 
+ \left( r^2+a^2\over r^2-2mr-a^2\right) \d r^2 
\nonumber\\
\fl &&
\qquad+ (r^2+a^2) (\d\theta^2 +\sin^2\theta\; \d \phi^2).
\end{eqnarray}
This is still Ricci flat, $R_{ab}=0$,
and one easily calculates
%\begin{equation}
%\fl
%R^{abcd} R_{abcd} = 
%- 48\; {(a^2-m^2)(r^2-a^2)([r^2+a^2]^2-[4ar]^2)-4ma^2r (3r^2-a^2)(r^2-3a^2) \over (r^2+a^2)^6}.
%\end{equation}
\begin{equation}
\fl
R^{abcd} R_{abcd} =  {a^2-m^2\over a^2} \; (R^{abcd} R_{abcd} )_{m=0} 
+  {192\; ma^2r (3r^2-a^2)(r^2-3a^2) \over (r^2+a^2)^6}.
\end{equation}
Much of the previous discussion follows with only minor modifications.

There is now a symmetry under the \emph{simultaneous} interchange of both $+r \longleftrightarrow -r$ and $+m \longleftrightarrow -m$. The curvature and surface area are still finite at $r=0$, and the natural range of the $r$ coordinate is again $r\in(-\infty,+\infty)$. The horizons are again coincident with the ergosurfaces but are now located at 
\begin{equation}
(r_H)_\pm = m \pm \sqrt{m^2+a^2}.
\end{equation}
Note that \emph{both} of these horizons should be taken into consideration; the second horizon $(r_H)_-$ cannot be summarily ignored as is done in references~\cite{Zhang, Chen, Ong}.
We now have
\begin{equation}
g_{tt} = -1 +{2m\over r} +  {2a^2\over r^2} + \O(1/r^3),
\end{equation}
indicating that the mass of the spacetime is indeed $m$ (at least as seen from the $r>0$ side, however it is effectively mass $-m$ as seen from the $r<0$ side). To check this observe that  the $\Gamma^r{}_{tt}$ Christoffel symbol is
\begin{equation}
\Gamma^r{}_{tt} = {(m[r^2-a^2]+2a^2r)(r^2-2mr-a^2)\over (r^2+a^2)^3 } \approx {m\over r^2} + \O(1/r^3) 
\end{equation} 
so the coordinate acceleration of a test particle dropped at rest is now inverse square. 
But the dominant inverse square term in the coordinate acceleration is now towards the nearest horizon for $m r >0$, and \emph{away} from the nearest horizon for $mr <0$. (To see some of the physical implications of negative asymptotic mass, at least in a spherically symmetric situation, consider~\cite{Visser:1995, Cramer:1994}.)

Other parts of the discussion change slightly, such as the precise location of the cigars of manifestly closed timelike circles, but the qualitative discussion remains the same.  In particular the causal structure remains pathological.

%------------------------------------------------------------------------------------------------------------------------------------------
%------------------------------------------------------------------------------------------------------------------------------------------
\section{Discussion}
%------------------------------------------------------------------------------------------------------------------------------------------
%------------------------------------------------------------------------------------------------------------------------------------------

As we have seen, the ``twisted'' black holes of~\cite{Zhang, Chen, Ong} are not really new, being minor variants of the quite well-known Taub--NUT spacetimes~\cite{Taub, NUT, Podolsky}. As mathematics, the Taub--NUT spacetimes are interesting examples of what can go wrong in general relativity if one blindly applies the field equations in isolation, without checking other features of the proposed spacetime~\cite{Visser:1995}. As astrophysics, the ``twisted'' black holes exhibit a number of extremely serious pathologies that make them unlikely candidates for real astrophysical black holes.

%------------------------------------------------------------------------------------------------------------------------------------------
%------------------------------------------------------------------------------------------------------------------------------------------
\ack
%%------------------------------------------------------------------------------------------------------------------------------------------
%%------------------------------------------------------------------------------------------------------------------------------------------

This research was supported by the Marsden Fund, 
through a grant administered by the Royal Society of New Zealand. 
FG is supported by the MacDiarmid Institute of the Victoria University of Wellington.
JS and SS are supported via Victoria University of Wellington PhD scholarships. 

\clearpage
%------------------------------------------------------------------------------------------------------------------------------------------
%------------------------------------------------------------------------------------------------------------------------------------------
%------------------------------------------------------------------------------------------------------------------------------------------
\section*{References}
%------------------------------------------------------------------------------------------------------------------------------------------
%------------------------------------------------------------------------------------------------------------------------------------------
%------------------------------------------------------------------------------------------------------------------------------------------
%------------------------------------------------------------------------------------------------------------------------------------------

%------------------------------------------------------------------------------------------------------------------------------------------
%------------------------------------------------------------------------------------------------------------------------------------------
\end{document}